# Communication and games in the online foreign language educational system. User behavior study.

Ilya V. Osipov [1*], Anna Y. Prasikova [1], Alex A. Volinsky [2*]

[1] i2istudy.com, Krišjāņa Barona Iela, 130 k-10, Rīga, Lv-1012, Latvija

[2] Department of Mechanical Engineering, University of South Florida, 4202 E. Fowler Ave., ENB118, Tampa FL 33620, USA

[*] Corresponding authors. Email: volinsky@usf.edu; phone: (813) 974-5658; fax: (813) 974-3539; Email: ilya@i2istudy.com (Ilya V. Osipov)

**ABSTRACT:**
The paper describes creation and development of the educational online communication platform for teaching and learning foreign languages. The system is based on the time bank principle, allowing users to teach others their native tongue along with taking foreign language lessons. The system is based on the WebRTC technology, allowing users to access synchronized teaching materials along with seeing and hearing each other. The platform is free for the users with implemented gamification mechanics to motivate them. It is based on the freemium model, where the main functions are provided free of charged with some premium features. The paper describes studies associated with user involvement in the learning/teaching process. The hypothesis whether two previously unfamiliar individuals could communicate with each other using a foreign language, based on the developed system algorithms, was tested. System virality, where new users are attracted by the existing users was also studied, along with user motivation for viral behavior. Relationships between monetization, virality and user involvement were also considered.

## INTRODUCTION:

The authors have created and developed online computer educational system for improving foreign language communication skills. The system allows communicating with native speakers in a foreign language. The system does not use Skype and does not offer interactive materials for self-study. Similar to Uber or AirBnB, where people offer their services without third parties, the system users teach each other foreign languages.

The logic is quite simple. If there are English-speaking people, who would want to learn Spanish, there will be Spanish-speaking people wanting to learn English. For this purpose it's important to create a common technological platform with educational materials, allowing users to communicate freely. There has to be a common place for an adequate number of people to communicate, which would contain simple and understandable materials allowing the student to learn from the native speaker using step-by-step instructions. Time banking and other game elements motivate users to utilize the system by switching roles. For example, native speaker can earn minutes by teaching the mother tongue, or spend minutes by learning a foreign language[1].

### System description and user behavior studies

Experiments were conducted using the users of the i2istudy system that utilized the service at their own will. To conduct the experiment the following advertising was displayed in Facebook and vk.com social networks: "Want to learn foreign languages for free, or teach your native language? Click here." Almost 40,000 users registered in the system from May through August 2014 as a result of the advertising. Most of the users wanted to learn English, 28,180, another 8,711 users wanted to learn Spanish, 1,028 Russian and 1,791 German. 14,943 users said that their native language is English, 20,673 Russian, 204 German, and 3,843 Spanish.

All registered users could find other users in the system present online, identified as students or teachers by pressing the corresponding "Learn X language" or "Teach Y language" buttons. This action sent requests to other selected users. Here, the words teacher and student refer to the user role in the system. Each user can assume both roles as a teacher of the native language and as a student learning a foreign language. The users select the corresponding languages during the registration process, and can be changed later.

The recipient of the teaching or learning request sees the user sending this request, along with the language and selected lesson level. The recipient can either accept or reject the request. If the request is accepted, the interface window opens, where the teacher and the student see and hear each other, and use synchronized teaching materials.

During the lesson the users not only see and hear each other, but jointly work with teaching materials divided into small portion (cards). Besides, the student can see prompts and translation in their understandable language. The teacher sees additional instructions, for example to ask the student to repeat words, etc. The system tracks the time spent learning and teaching. At the end of the lesson each participant rates the other party's corresponding quality of teaching or learning.

Experiments were conducted using an online educational system with 40,000 registered users and 1,000-1,500 daily active users. The research objective was to test the assumption that two unfamiliar people could communicate and learn a foreign language together within the system framework. To test whether this hypothesis is even feasible, when specifically unmotivated people, without prior training could assume the appropriate role of a "student" or a "teacher", find and invite another user online to teach or to learn a foreign language. The system interface contains the tools for finding other users currently available online (Figure 1). The lessons are following pre-defined scenarios using face-to-face communication, realized using WebRTC technology[2,3].

*[Place Figure 1 here]*

**New user attraction methods**

Initial users interested in practicing foreign languages registered as a result of an advertising placed in the Facebook social network. The advertising suggested registering in the system to learn foreign languages for free in exchange for teaching native language. The ads were displayed in English and Spanish-speaking countries, Germany and Russia. Besides, some of the users were invited by their friends (new users invited by the existing users). As a result, 39,729 total users registered in the system, including 28,180 users who wanted to learn English, 8,711 Spanish, 1,791 German and 1,028 Russian languages. Wherein, 14,943 users said that their native language is English, 20,673 Russian, 3,843 Spanish and 204 German.

If the "teacher" accepts the "student" invitation and visa versa, a live audio-video session is established. Both users can hear and see each other, along with the synchronized lesson materials. These materials include corresponding individual prompts for the participants in their chosen language. The system also tracks the lesson time for the billing purposes in terms of the game currency or real money[4,5].

About 20% of the registered users participated in the experiments. The rest were either intimidated to talk with strangers online, or did not understand how the system works and decided not to spend their time. Some users could not configure their hardware (microphone and web camera) needed for the online communication session, or their browser did not support the WebRTC protocol used by the system. Table 1 shows the absolute number and percentage of the newly registered users for each month that made a call

**RESULTS:**

As a result of the test, it was established that two unknown people, who met in the developed online interactive system for the first time, could carry on a conversation in a foreign language and effectively help each other. Moreover, part of the users did not speak a common language, and communicated by using system prompts displayed in their respective native language. The average successful connection time was about 12 minutes (189,207 minutes or 3153 man-hours) divided by the 15,842 total successful connections. Any type of connection termination was accounted for, including finishing the lesson, participants terminating the connection within the system, or by simply closing the browser. Table 2 shows the number of successful connections for every month and the total connection time in minutes.

Regardless of the fact that the average connection time was not very long, this experiment shows that two unprepared participants, previously unknown to each other, can carry on a conversation for an extended period of time using the developed system. If the lessons duration was not too long, the users could finish them. Moreover, the average connection time continued to increase with the number of active users, reaching over 14 minutes by the end of the experiment in August 2014.

**Retention and viral user attraction. K-factor**

Besides the main cycle of the freemium product utilized by the users, other cycles can be identified. The retention cycle involves users leaving and coming back to the application. The viral cycle involves viral mechanics to attract new users by existing users inviting their friends. The monetization cycle involves utilizing system premium paid services, allowing the system to make money[6].

It's interesting that these cycles can be partially antagonistic, since they compete for the user attention, which is a limited resource. They can also act cumulatively by helping each other. One of the ways to develop freemium products is to attract the maximum number of users, followed by monetization. Moreover, new user acquisition is achieved with the help of the existing users. The user base growth is called viral in this case.

Targeted metrics of how new users are attracted by the existing users is described in terms of the K-factor, also called the viral factor[7]:

$$K_{factor} = AiPU \cdot IPi \qquad (1),$$

Here, $IPi = IU/i$ the ratio of the newly registered invited users, $IU$, and the total number of sent invitations, $i$. $AiPU = i/U$ is the ratio of the total number of invitation sent by the users, $i$, and the

Page 4 of 20

total number if users, *U*. Note that the total number of sent invitations, *i*, cancels out, and the K-factor = $U*IU$. However, without the invitations the system could not grow virally, which is why equation (1) reflects the practical meaning of the viral K-factor.

To practically track the system dynamics the author calculate the local K-factor for a given period of time, one week in this case. It's important to emphasize that the local K-factor reflects the percentage of the newly attracted users in relation to the all existing users. The local K-factor value can be much different from the global K-factor, which reflects the average number of new users attracted by the existing users over the lifetime of the project. Moreover, for the purity of the experiment, only active users were considered. Using the local weekly K-factor allows tracing the system dynamics affected by certain changes and innovations. Additionally, the K-retention parameter tracks how the audience is retained by the project. The ideal K-factor is 1. In this case, the project ideally retains the audience without loosing a single user[6]:

$$K_{retention} = \frac{dU - dNU}{dU_{-1}} \qquad (2)$$

Here, *dU* is the daily audience for a given day, *dU*$_{-1}$ is the audience in the previous day, and *dNU* are the newly registered users in this timeframe. For example, if K-factor is 20%, and the K-retention is 90% (meaning that 9 out of 10 users visit the service the next day), then the growth K-factor is 0.2+0.9=1.1. This system will continue to grow by 10% of its daily audience day by day[7].

*[Place Figure 2 here]*

For example, Figure 2 shows the user growth with 1 month increments, assuming 1000 starting users, 20% monthly K-factor, 85% K-retention factor, extrapolated over 36 months.
The sum of the K-factor and the K-retention factor is called the K-growth. If K-growth is larger than 1, geometric growth progression would result. Alternatively, if K-growth is less than 1, the system looses users[8-10].

**Monetization as system control**

Regardless of the fact that the system was initially designed as predominantly free, at a certain point users were given an option to purchase certain features. Internal system currency is time in minutes gained by teaching and spent by learning. Users were given an option to purchase minutes in the system using real money (Figure 3). This feature to purchase minutes was added to the user interface as a control action[11,12].

*[Place Figure 3 here]*

Let's see how the users reacted to the made change. The authors expected that 1-2% of the users would utilize the new option to purchase minutes. This seems to be realistic, as typical freemium products have 2-10% paying premium customers[6,13].

**Relationships between monetization and virality**

Schematics in Figure 4 shows how users could earn or spend minutes in the system by A-inviting friends, B-spending minutes as a student, C-by teaching other users and D by purchasing minutes, which was the newly added feature.

*[Place Figure 4 here]*

The actual user interface implementation to purchase minutes is shown in Figure 3. As mentioned earlier, the user had the free will to either use this feature or not. However, the developer's expectations have not been met, since only two users purchase system minutes since July 28$^{th}$, 2014. Thus, only 0.0083% of the users got involved in the offered monetization, which is unacceptable[14].

However, this new feature has demonstrated the value of the minutes in the system to the rest of the users, which caused an unexpected growth of user involvement and virality. As a result, the number of users teaching their native language increased. Prior to this monetization feature implementation 47% of the active users taught in the system, while 55% of the users taught after they were exposed to the value of the system minutes. The K-factor growth dynamics was even more interesting. Figure 5 shows the weekly K-factor before and after the implemented monetization feature.

*[Place Figure 5 here]*

As seen in Figure 5, the weekly K-factor increased when the new function was added, and continued to grow until September 1$^{st}$, 2014, when the system was closed for renovation. Before July 28$^{th}$, the average weekly K-factor was 2.2%, and increased to 3.8% after monetization.

To test the hypothesis of the monetization effect on the K-factor, the p-value was calculated. The null hypothesis assumed that the K-factor did not change as a result of the monetization. The K-factor data before monetization was used for the expected value. The actual K-factor data was taken after monetization. The calculated p-value was 0.01%, thus the authors have rejected the hypothesis that the K-factor did not change as a result of monetization[14,15].

Thus, it is clear that trying to affect the user behavior to motivate them by the paid functions resulted in an unexpected opposite effect. The users started to use alternative ways to earn minutes in the system, but not real money.

**System interface to motivate user behavior**

In order to increase the virality and attracting new users, the existing users were asked to invite their friends. This is how the corresponding system interface looked like (Figure 6).

*[Place Figure 6 here]*

This simply did not work, as the number of newly attracted users was minimal, corresponding with the K-factor of 1-1.15% in May 2014. Thus, the message was changed to ask users for their help to grow the project (Figure 7 a).

*[Place Figure 7 here]*

The loyal system users were addressed, as the message was displayed in the internal system interface after the users logged into the system. This message had a much stronger effect, as 26% of the users clicked the "Help the project" button in Figure 7a. As a result, the K-factor grew by 1%. The authors decided to strengthen the effect by removing the closing window cross in the upper-right corner of the window (Figure 7b). Thus, the user could only reject help consciously. It was impossible to close the window. As a result, 73% of the users invited their friends, and the K-factor grew to 4%[16]. These very simple motivating aids increased the key parameters of the project in terms of its growth[17,18].

**DISCUSSION:**

The ideas of cooperative learning in the form of a game are accepted by the users quite well. Regardless of the existing stereotype that only "professionals" can provide quality foreign language education, our system users learned together and helped each other without the aid of "professional" foreign language teachers, just based on the developed system, which contains pre-defined materials along with the prompts understandable by each user in their native language. In this case, similar to the interactive materials, the professional foreign language teacher role is to develop learning materials. The users can learn from these materials with other native speaker users, based on the social interaction effect. Based on the conducted experiments, the existing users not only spend the time in the system learning foreign languages, but also attract their friends.

**Conclusions**

Based on the conducted experiment, the average connection time increased to almost 15 min with many users involved in system integration. Some of them became daily active users. Users were motivated to invite their friends using simple methods described in the paper. As a result, the K-factor has crossed over the 4% mark. The authors suppose that this type of learning could partially replace individual tutoring to improve one's foreign language oral communication skills.

Figure Legends:

Figure 1: The system interface with the tools to find other users currently available online.

Figure 2: Hypothetical user growth with 1 month increments, assuming 1000 starting users, 20% monthly K-factor, 85% K-retention factor, extrapolated over 36 months.

Figure 3: System interface with the new option to purchase minutes.

Figure 4: Schematics showing how users can earn or spend minutes in the system.

Figure 5: The weekly K-factor showing that it doubled after added monetization.

Figure 6: System interface offering 30 minutes for each invited friend.

Figure 7: a) Message asking users for their help to grow the project with the red window-closing cross in the right-upper corner; b) Similar interface with the window-closing cross removed and the added "I would not like to help" button.

Table 1: The number and percentage of the newly registered users for each month that made a call.

Table 2: The number of successful connections every month and the total connection time in minutes.

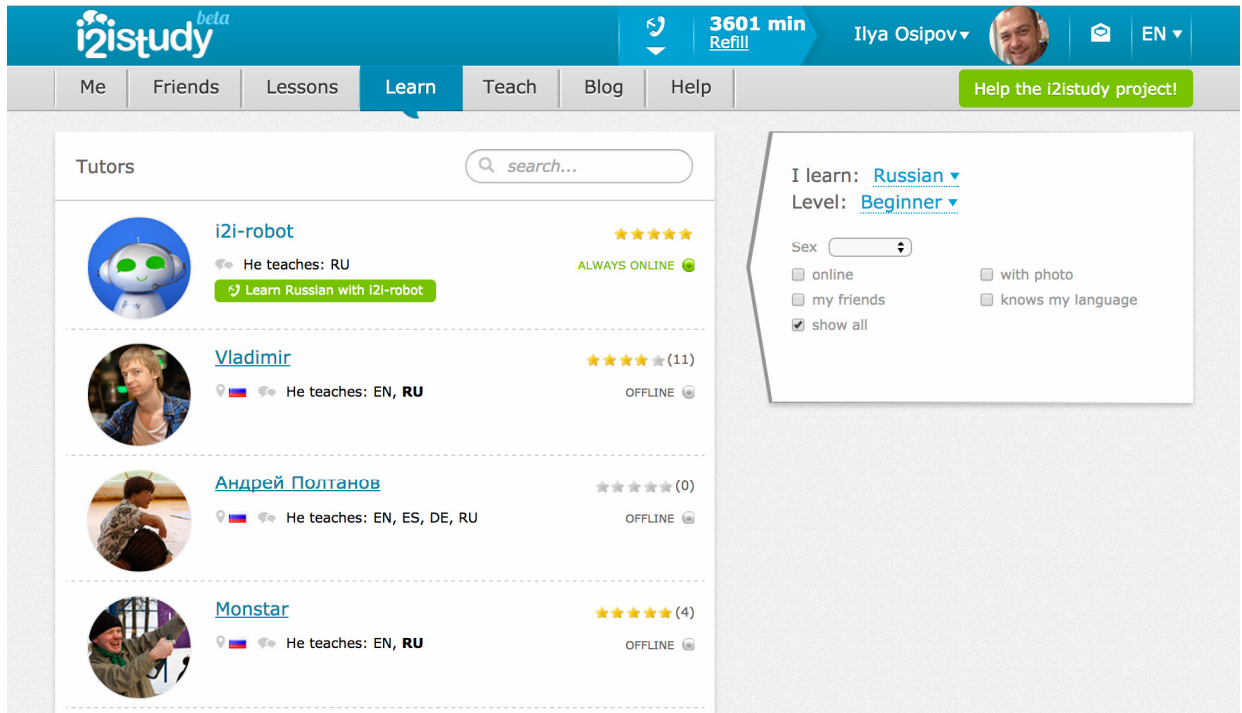

**Figure 1: The system interface with the tools to find other users currently available online.**

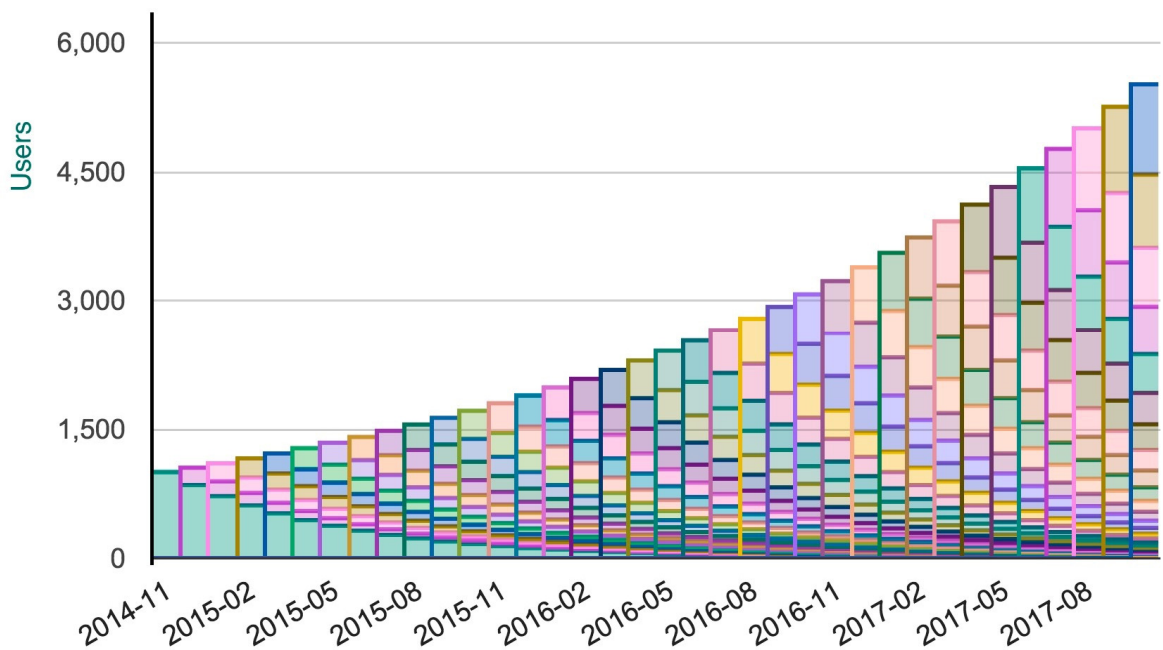

**Figure 2: Hypothetical user growth with 1 month increments, assuming 1000 starting users, 20% monthly K-factor, 85% K-retention factor, extrapolated over 36 months.**

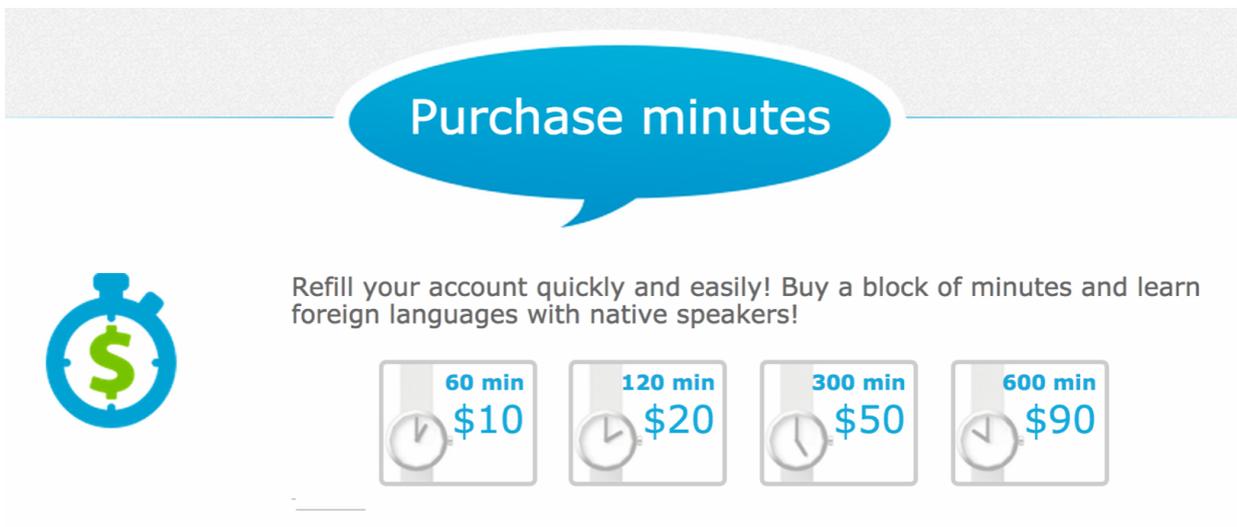

**Figure 3: System interface with the new option to purchase minutes.**

**Figure 4: Schematics showing how users can earn or spend minutes in the system.**

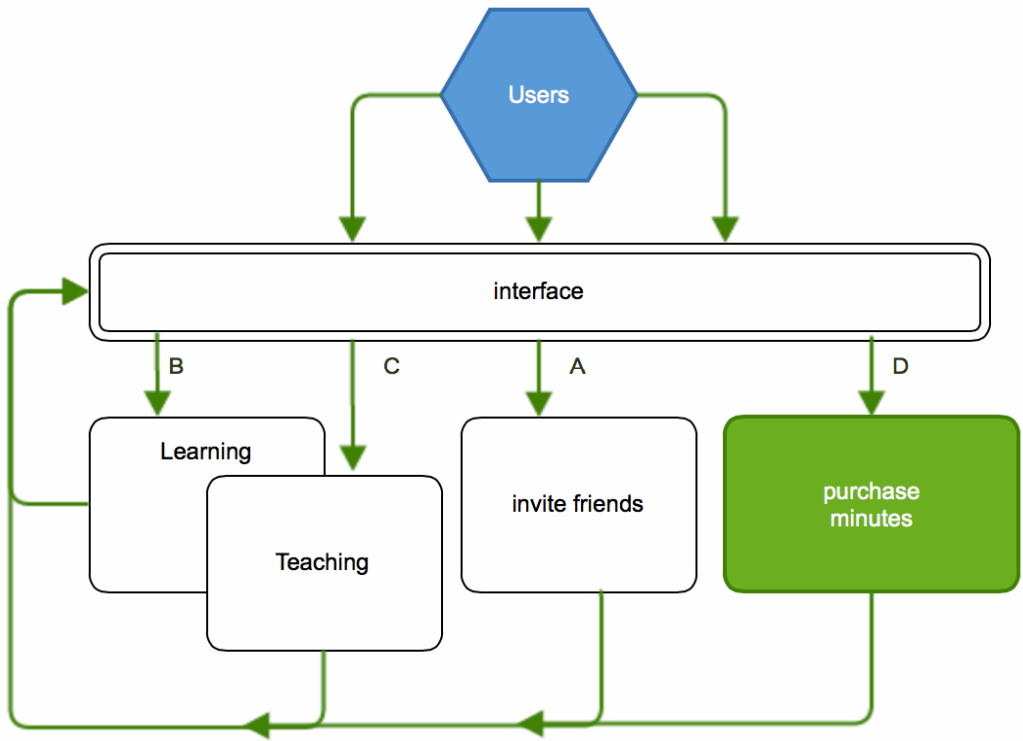

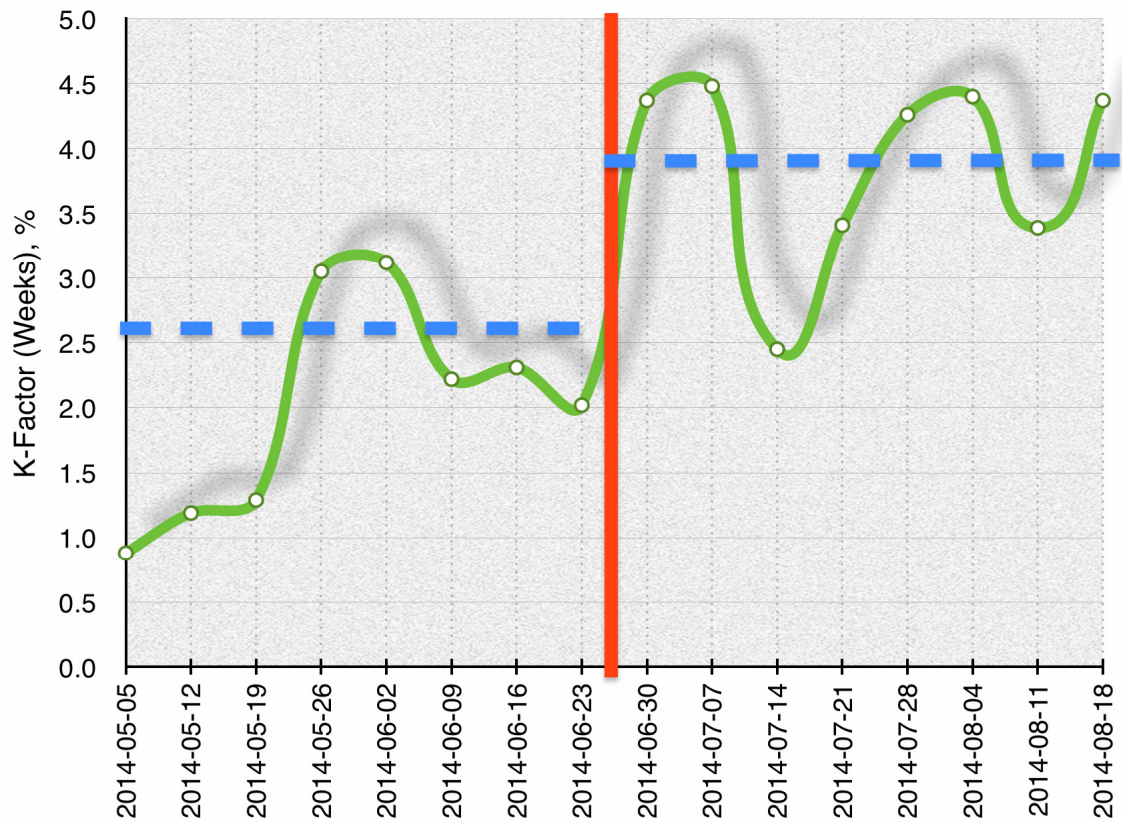

**Figure 5: The weekly K-factor showing that it doubled after added monetization.**

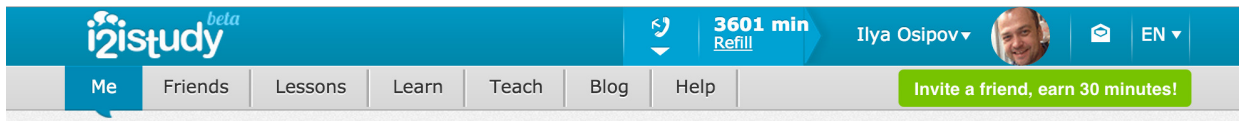

**Figure 6: System interface offering 30 minutes for each invited friend.**

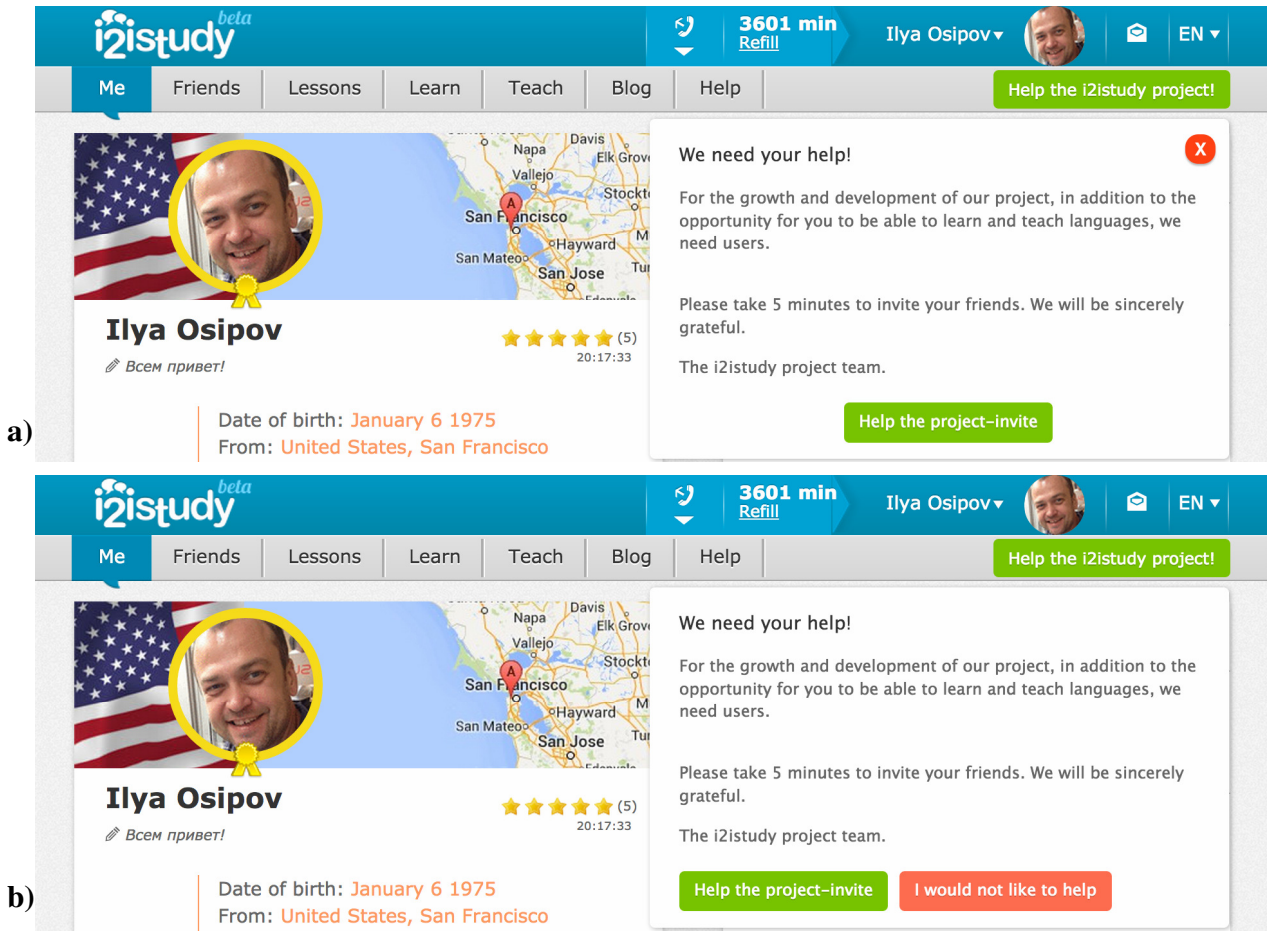

**Figure 7: a) Message asking users for their help to grow the project with the red window-closing cross in the right-upper corner; b) Similar interface with the window-closing cross removed and the added "I would not like to help" button.**

**Table 1: The number and percentage of the newly registered users for each month that made a call.**

**Involvement**

| Month | 01.12 - 31.12 | 01.01 - 31.01 | 01.02 - 28.02 | 01.03 - 31.03 | 01.04 - 30.04 | 01.05 - 31.05 | 01.06 - 30.06 | 01.07 - 31.07 | 01.08 - 31.08 |
|---|---|---|---|---|---|---|---|---|---|
| Number of just registered users who made a call in period | 93 | 734 | 61 | 15 | 251 | 1026 | 2037 | 2072 | 722 |
| Percent of just reg. users who made a call in period | 7 | 16 | 22 | 12 | 22 | 26 | 25 | 18 | 15 |

**Table 2: The number of successful connections every month and the total connection time in minutes.**

**Real connects**

| Month | 01.12 - 31.12 | 01.01 - 31.01 | 01.02 - 28.02 | 01.03 - 31.03 | 01.04 - 30.04 | 01.05 - 31.05 | 01.06 - 30.06 | 01.07 - 31.07 | 01.08 - 31.08 |
|---|---|---|---|---|---|---|---|---|---|
| Successful connects duration, min | 151 | 10835 | 5021 | 3645 | 6202 | 37037 | 47140 | 54974 | 38202 |
| Number of successful connects | 19 | 1228 | 492 | 131 | 587 | 3093 | 3868 | 3763 | 2661 |